D'Acci L., 2012. arXiv
Physics and Society (physics.soc-ph) General Finance (q-fin.GN)# Isobenefit Lines, Breaking Point of equal attraction, Uniformity Benefit, Variety Value and Proximity Value, Preference Gap Gain.

**Luca D'Acci**
lucadacci@gmail.com*Isobenefit Lines* can offer a certain range of applicability in Location Theory and Gravitational Models for Urban and Geography Economics, in positional decision processes made by citizens, and, last but not least, in land value and property market theories and analysis. The value of a land, or a property, in a generic *k* point, is, *ceteris paribus*, the mirror of the quality, attractiveness, benefit characterizing *k*. *Preference Gap Gain* (PGG) of a person, is the difference between his *Personal Isobenefit Lines* and that of the majority of people. In monetary terms, when buying or renting a property, it can become an economic gain or vice versa, and PGG localizes and quantifies this gain.## 1. Isobenefit Lines definition

The *Isobenefit Lines* (D'Acci 2006, 2007, 2009a,b, 2012a,b,c,d), join the urban points with a same level of benefit given from urban amenities. Considering amenities the urban attractions such as parks, pedestrian streets, nice squares, pleasant shopping areas, etc.
The benefit of a generic point (*k*) in the city received from a generic amenity *i* with a level *A* of attractiveness, is given by (D'Acci 2009a,b, 2012a,b,c,d):

$$B_{i,k} = \frac{A_i}{1 + \frac{d_{i-k}}{E}}$$

(1)

Where *E* is a coefficient of "Efficiency of moving" that depends from the cost/comfort/speed of reaching the amenity. Fig. 1 shows the result of equation 1.

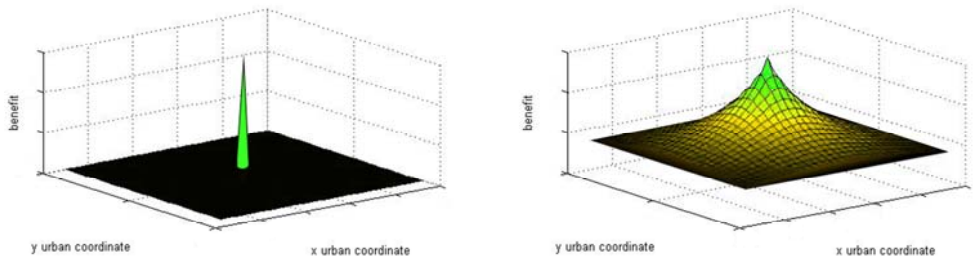

**Fig. 1** Example of Spatial Benefit (right) from an amenity (left)

*E* should not be too high (Fig. 2), because it is suggested a function shape where the gradient of the benefit, against the distance, varies significantly (like a parabolic or hyperbolic shape). If not, following the equation 1, a point between two amenities could result with a analogous benefit than a point in front of one of the two amenities.

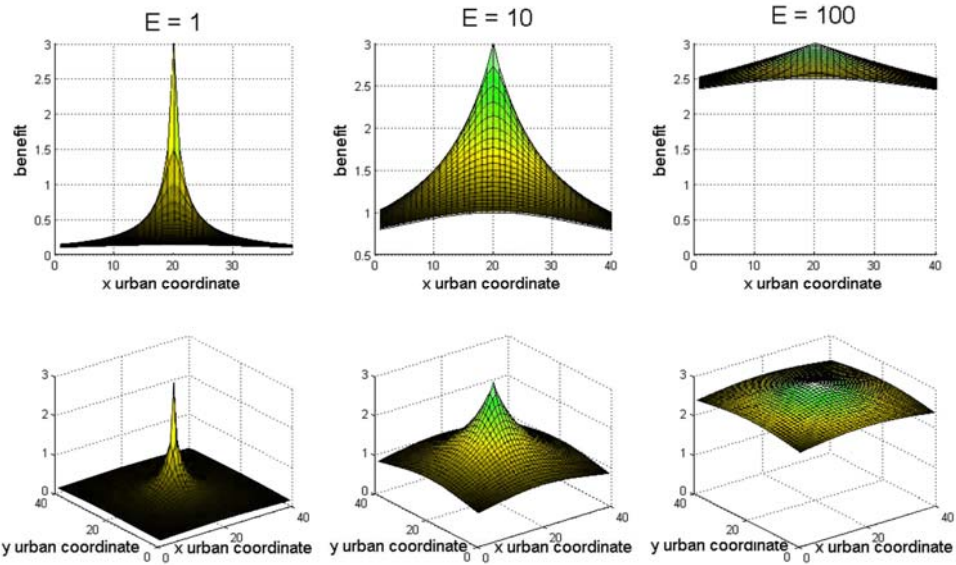

**Fig. 2** Examples of different values of the Moving Efficiency coefficient (*E*) with Ai=3 by equation 1

We could also choose other kinds of function shapes (Fig. 3) such as (Mossello 1990):

$$B_{i,k} = A_i \cdot e^{-E \cdot (d_{i-k})^2}$$

(2)

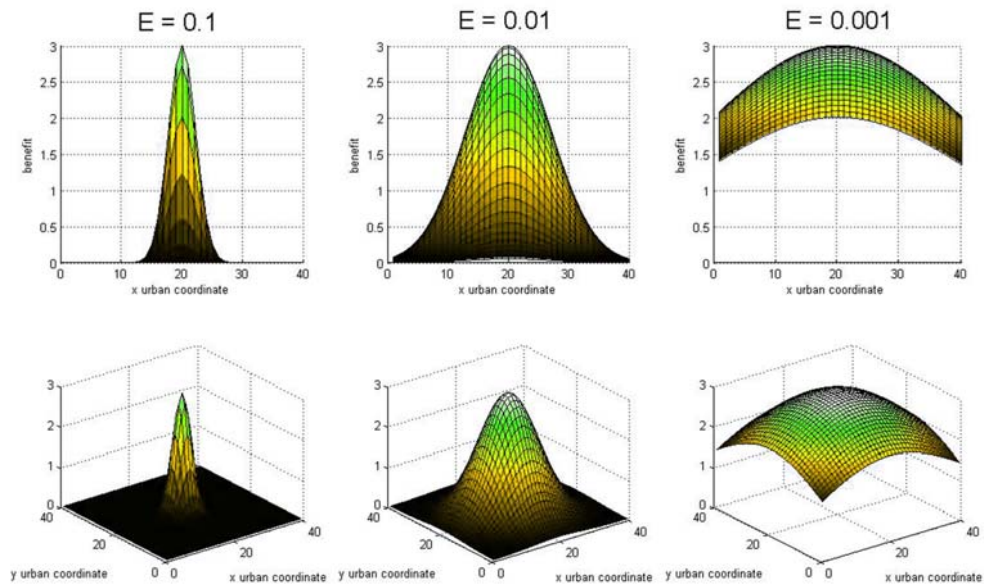

**Fig. 3** Example of Gaussian relation between accessibility and distance

Or (Fig. 4), by deleting the squared in equation 2 (similarly to Wu & Plantinga , 2003, however with other meaning and in other contexts):

$$B_{i,k} = A_i \cdot e^{-E \cdot (d_{i-k})}$$

(3)

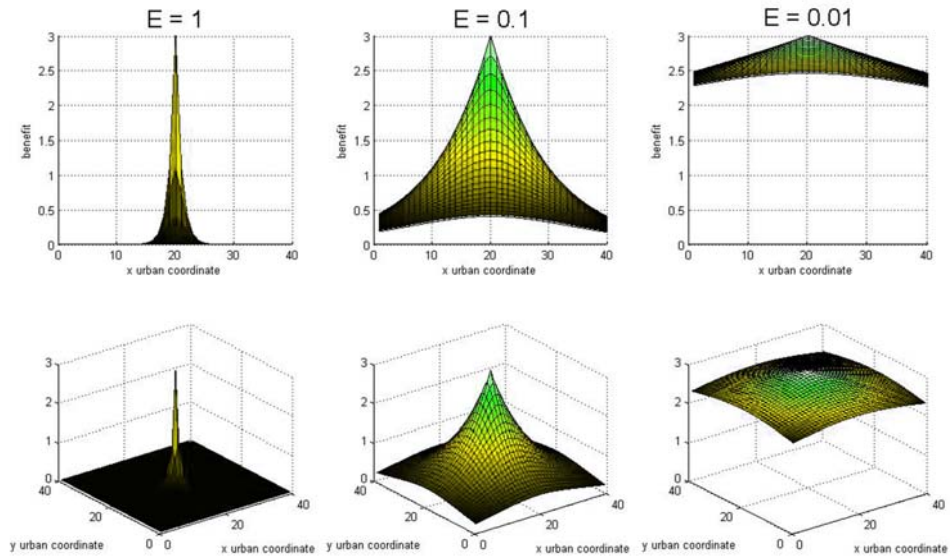

**Fig. 4** Example of exponential relation between accessibility and distance

We can also refer to concepts and models well-known in geography such as "Accessibility Potential" and Gravitational Models, or to prior methods developed by location theorists (Malthus, von Thünen, Christaller, Losh, Alonso, Muth, Mills, Smith, Isard, Moses, Hotelling…), the weberian Isocost Lines and their derivate concept of Isodapanes (Geertman and Van Eck, J. R., 1995), etc.

Independently from the equation chosen, the *Isobenefit Lines* are the lines that join the urban points with the same level of $B$ (Fig. 5).

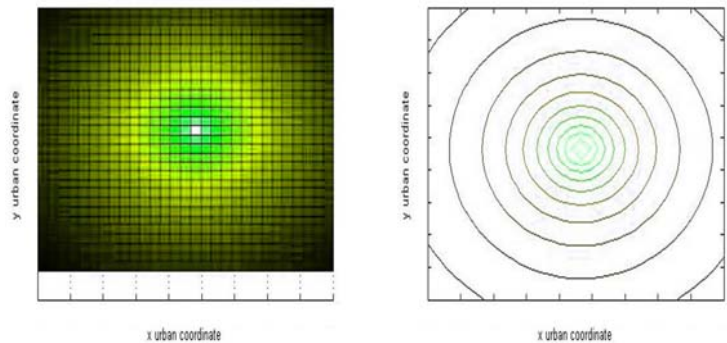

**Fig. 5** Isobenefit Lines

For the 'non-attractiveness' of the city (busy streets, abandoned factories, cemetery, etc.), we introduce an *A* with a negative value (Fig. 6).

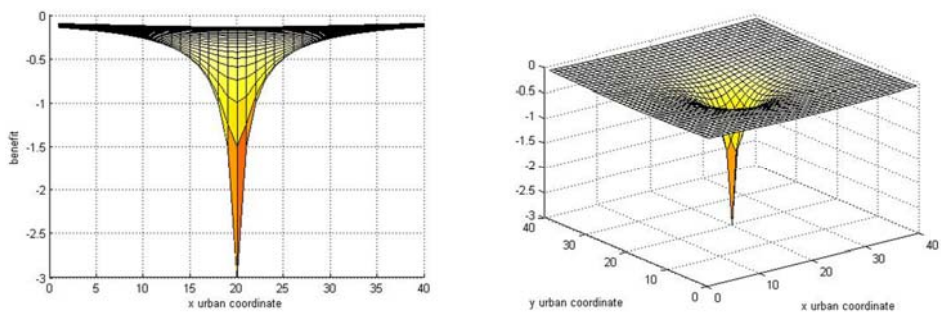

**Fig. 6** 'Non-attractiveness'

Taking into account the entire attractions present in a city, the benefit of the *k* urban point is the sum of the benefit given by each attraction in the city, and it depends on the distance, level, number and reciprocal positions of attractions:

$$B_k = \sum_{i=1}^{n} B_{i,k} = \sum_{i=1}^{n} \frac{A_i}{1 + \frac{d_{i-k}}{E}}$$

(4)

to quantify the uniformity of the spatial distribution of the attractions *effects* on the city, by the Isobenefit Lines, we can use the following indicator (the *Uniformity Coefficient*, D'Acci 2012a,b,c,d, 2009a,b):

$$U = 1 - \sqrt{\frac{\sum_{k=1}^{m}\left(B_k - \frac{1}{m}\sum_{k=1}^{m} B_k\right)^2}{m}} \bigg/ \frac{1}{m}\sum_{k=1}^{m} B_k$$

(5)

$U$ is a number less or equal to 1 (maximum uniformity of the benefit effect). We should separately consider amenities ($A>0$), and disamenities ($A<0$), in order to quantify $U$.

**2. Variety Value and Proximity Value**

$U$ is 1 minus the variation coefficient (namely the ratio between standard deviation and mean) of the benefit of each urban point. It does not quantify the uniformity of the amenities location throughout the city, but the uniformity of the positional advantage of each urban point. If we compare different scenarios, $U$ should be read together with other indicators such as the total, medium, maximum and minimum value of $B$. For example, we imagine a radiocentric city with all its attractions distributed on the external circular crown (scenario 1), or throughout all the city area (scenario 2). It could (depending on $E$) even result $U1>U2$. However, we will notice that the level of $B$ is higher in each point in scenario 2 rather than in 1. $U$ could result higher in 1 than in 2 because the central part has always a positional advantage due to its geometrical position, even if it does not have any amenity in front of it (scenario 1). In scenario 1, it has an advantage of "variety"; it can get all the amenities with the same effort (distance). If we put amenities uniformly covering all the city area (scenario 2), the positional advantage of the centre becomes even higher in comparison with the periphery ($U1>U2$). In equations 2, 5 and subsequent, the higher $E$, the higher this distortion of the lecture of $U$ (and B) could be. For this are suggested low values of $E$.
The higher $E$ the more the equation 'weighs' the 'variety' advantage to enjoy numerous amenities, rather than the advantage of the proximity of one amenity. I defines the first advantage as *Variety Value*, the second as *Proximity Value*.

**3. Breaking point of equal attraction**

Huff (1963) provides the probability (P) of a customer (C) living in a place (i), to travel to a particular facility (j) distant $d_{ij}$, considering all the other n facilities available:

$$P(C_{ij}) = \frac{S_j / d_{ij}}{\sum_{k=1}^{n} S_k / d_{ik}}$$

(6)

By substituting facility with amenity – urban attractions – and customer with citizen, equation 6 can easily be read in another point of view. The same can be done for several gravitational models available. For example, the Isobenefit lines can offer a comparison with the Breaking point of Reill's law of Retail Gravitation (1931). It describes the breaking point of the boundary of equal attraction. Reading Reill's equation from the point of view of urban amenities:

$$Br_{1,2} = \frac{d_{1,2}}{1 + \sqrt{A_1/A_2}}$$

(7)

Where *Br* is the breaking point between the urban amenity 1 and 2, $d_{1,2}$ is their reciprocal distance, and *A* their attractiveness. This can be compared with the point, *Br*, showed by the Isobenefit lines (from equation 1, 4) in Fig. 7: the minimum value of benefit between the two amenities. It is the point at which a marble placed in the Isobenefit surface settles.

The breaking point can be personal: one can prefer a closest amenity even if less attractive. Change of preference also happens for a same person at a different stage/age of his life.

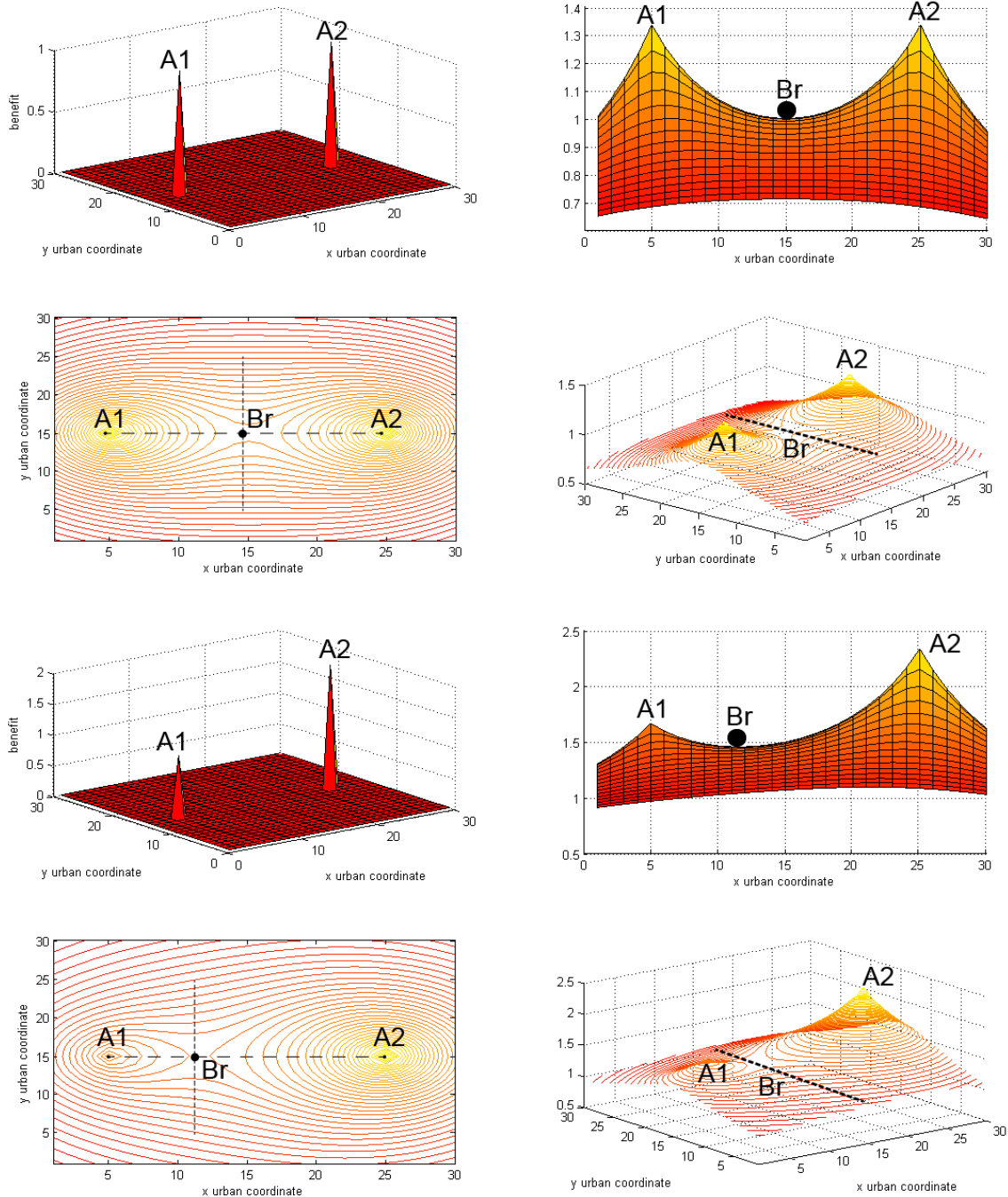

**Fig. 7** Example of breaking points

## 4. Personal Isobenefit Lines

We can also build *Personal Isobenefit Lines* matching the preferences of each citizen. *Isobenefit Lines*, can vary among people, and ages. For example, one could prefer the variety offered from the availability to access more than one amenity (and with a greater level of attractiveness), even paying the cost of not enjoying living very close to any amenity (although with a low attractiveness).

If we build Personal Isobenefit Lines, $E$ will also be influenced by the personal propensity to move, or the personal preference for the Variety Value (directly proportional to $E$ in eq. 1, 4) rather than for the Proximity Value (inversely proportional to $E$).

Isobenefit Lines can also be personalized because each person can feel different levels of attractiveness ($A$ in equation 1, 4) for a same amenity.

We should numerically judge how much an attraction can satisfy its pretension to be an 'Attraction' for the majority of citizens. It could be sensible to judge $A$ by referring to the usual, average number of citizens (not tourists) using the attraction and by then comparing each amenity with the best place/s in the city and with the neutral ones. In a similar way, but in different contest and aim, urban economists have often been interested in using population levels as a measure of urban success. High levels of population "tell us that people are voting with their feet to move to a particular place" (Glaeser 2008). There is no doubt about the relativity, and then the validity, of our own preferences also if divergent from other people, or even from the average peoples preferences: that which for a person can be a wonderful attraction, i.e. a shopping mall, for another can be a boring, consumerist place. Idem for the judgment of amenities such as parks, historical areas, and so forth.

## 5. Preference Gap Gain: Property value as spatial benefit mirror

If, *ceteris paribus*, the average price of properties in two different urban areas of the city (D'Acci 2012e), are the same, it seems that citizens (customers) showed to appreciate both areas in the same way; or that a part of them preferred one area, and the other part the other area, but both in the same way.

If, *ceteris paribus*, the average price of properties in two different urban areas are different, those prices are usually expected to reflect the preference/needs expressed by citizens.

This is valid as long as there is an adequate volume of trade of houses in the areas. In fact, to fix a price, it is enough that there is one person (not the majority) willing to pay that price. However this also depends from the time that the seller is willing to wait for selling, and from the ratio between the number of sellers and buyers from that kind of property/land.

Comparing the offer price, that everybody can see in any real estate agency website, of a similar land/property but in different places such as a nice city centre, an unattractive peripherical area, a touristical mountain locality, a countryside area without services or comfortable streets, etc., we could notice differences. Those differences reflect in part the preferences/needs shown from people and their costs/advantages by living there, i.e. commuting, amenities, average income/job offer in the area, etc.

That citizen whose preference diverges from the average preference/needs, when buying or renting a property, could have an economical advantage, or a disadvantage depending on the direction of the divergence.

I call *Preference Gap Gain* (*PGG*) this advantage/disadvantage. Advantage, i.e, is when one prefers an area in which usually nobody would like to live (a quite inaccessible point, distant from any services, centralities, non touristic, etc.); disadvantage is when this person is obligated (for some reason), to live in an area that he does not like but that everyone loves. In the first case he could pays less for something which for him has a great value, vice versa in the second case. This is the well-known surplus of the customer (Dupuit 1844, Marshall 1890).

Those examples are valid if we consider the income and the commuting cost as a constant and not depending from the area in which the citizen chose/must live. Income in big and/or expansive cities is usually higher than in small/cheap cities, and this is compensated (spatial equilibrium) from the higher housing costs, general living costs and/or some disadvantage (Glaeser 2008). Therefore the examples mentioned can refer to citizens choosing a place inside a same city (that means having the same job), or across cities/places (countryside, cities, villages…) if their job does not change by moving and does not require commuting change (i.e. independent worker, such as a writer, web designer working online, etcetera).

Under those conditions, *PGG* of the person x, can be visualized/measured by overlapping/subtracting the Isobenefit Lines of x and the Isobenefit Lines of the majority of people.

**Acknowledgments**

I would like to thank Professor Pietro Terna for his advise

D'Acci L., 2012. arXiv
Physics and Society (physics.soc-ph) General Finance (q-fin.GN)